\documentstyle[preprint,aps]{revtex}

\begin{document}
\draft
\preprint{\begin{tabular}{l}SNUTP 94-19\\
                            March, 1994\\ \ \end{tabular}}
\title{Cosmological Constant and Soft Terms in Supergravity}
\author{Kiwoon Choi,$^{(a)}$ Jihn E. Kim$^{(b)}$
and Hans Peter Nilles$^{(c,d)}$}

\address{\sl $^{(a)}$ Department of Physics,
Korea Advanced Institute of Science and Technology \\
373-1 Kusong-dong, Yusong-Gu, Taejon 305-701, Korea\\
$^{(b)}$Department of Physics and Center for Theoretical Physics,
 Seoul National University\\ Seoul 151-742, Korea\\
$^{(c)}$Physik Department, Technische Universit$\ddot {\sl a}$t
M$\ddot {\sl u}$nchen\\
D-85747 Garching, Germany\\
$^{(d)}$Max-Planck-Institut f$\ddot {\sl u}$r Physik,
Werner-Heisenberg-Institut\\
D-80805 M$\ddot {\sl u}$nchen, Germany}

\def\be{\begin{equation}}
\def\ee{\end{equation}}

\maketitle
\begin{abstract}
Some of the soft SUSY breaking parameters in hidden sector supergravity model
depend  on the expectation value of
the hidden sector scalar potential, $\langle V_h\rangle$,
whose tree level value is equal to the tree level cosmological
constant. The current practice of calculating soft parameters
assumes that $\langle V_h\rangle=0$.
Quantum correction to the cosmological constant
can differ from the correction  to $\langle V_h\rangle$
by an amount of order $m_{3/2}^2M_{Pl}^2/8\pi$.
This implies that, for the vanishing cosmological constant,
the $\langle V_h\rangle$-dependent parts of soft terms
can be sizable, and
hence the supergravity phenomenology should be accordingly modified.
\end{abstract}

\newpage
Presently supersymmetry (SUSY)  is widely believed to be a
leading candidate for physics beyond the standard model.
This is largely due  to the fact that SUSY provides
the only known perturbative solution to the problem of quadratic
divergence in the Higgs boson mass.
Phenomenologically
viable supersymmetric models contain soft supersymmetry
breaking terms which are presumed to be induced by
supergravity interactions
in underlying $N=1$ supergravity theories$\cite{nilles}$.
Most of supergravity phenomenology then depend upon
the
nature of soft terms in the
resulting gobally supersymmetric effective theory.

An interesting feature of supergravity models
is that  many of the coefficients of soft terms are
calculable at tree approximation.
Furthermore, in some cases the model predicts certain
tree level relations
among the soft parameters renormalized at
the Planck scale $M_{Pl}$.
For instance, in the dilaton-dominated
SUSY breaking scenario in string theory$\cite{louis,ibanez}$, one finds
$A_{ijk}=-\sqrt{3}m_0\lambda_{ijk}$
where $A_{ijk}$ denotes a generic trilinear scalar coefficient,
$\lambda_{ijk}$ is the associated Yukawa coupling, and $m_0$
is the soft scalar mass.

For physical applications of any tree level result,
one needs to  take into account quantum corrections.
Ordinary renormalizable gauge and matter interactions would lead to
one loop corrections proportional to
$\frac{g^2}{8\pi^2}\ln(\Lambda^2/\mu^2)$
where  $g$ denotes a gauge or Yukawa coupling
constant,  $\mu$ is a scale around the weak scale $m_W$
where the loop graph is being evaluated,
and $\Lambda$ is the momentum cutoff above which the validity
of four-dimensional $N=1$ supergravity theory breaks down.
For $g^2$ of order unity and $\Lambda$ around $M_{Pl}$,
these corrections become important due to a large logarithm.
A nice feature of this type of corrections  is that they
do not depend strongly
on the cutoff $\Lambda$ and thus are  {\it calculable} within
the supergravity model by the aid of renormalization group analysis.

Besides the above type of logarithmic corrections, there
are other types of corrections which depend strongly
on $\Lambda$ and
thus whose precise magnitudes are {\it not} calculable
 within the supergravity model$\cite{gaume}$.
For instance, one loop graphs induced by nonrenormalizable gravitational
interactions would give power-law divergent corrections
which are proportional to
$(\kappa_0\Lambda)^2/8\pi^2$
where $\kappa_0=\sqrt{8\pi}/(M_{Pl})_{\rm bare}$ denotes the dimensionful
{\it bare} gravitational coupling constant.
To estimate the size of
such corrections,
one needs to determine the dimensionless coupling
`$\kappa_0\Lambda$', which  requires an information of
the theory underlying the supergravity model at energy
scales above $\Lambda$.
For instance, for a  supergravity model
which corresponds to the low energy limit of heterotic string theory,
it is most natural
to set $\Lambda$ to the string scale
 $M_{\rm string}$.
We then have
 $\kappa_0\Lambda=\kappa_0 M_{\rm string}=g_{\rm GUT}$
where $g_{\rm GUT}$ is the unified gauge coupling constant.
Inspired by this observation, throughout this paper,
we will assume $\kappa_0\Lambda=g_{\rm GUT}$ whenever we consider
a power-law divergent quantum effect.

For $\kappa_0\Lambda=g_{\rm GUT}$,
generic power-law divergent corrections would be expected to be  of
order ${\alpha_{\rm GUT}}/{2\pi}$
and thus not so significant.
However, there is one important exception for this naive
expectation.
At one loop order, the cosmological constant
(the vacuum energy density)  receives
a quadratically-divergent zero point
energy contribution$\cite{zumino}$.
If the model contains  $N$ chiral multiplets with
the boson mass $m_B$ much larger than the fermion
mass  $m_F$,
the cosmological constant receives a contribution:
\be
N\int {d^3k\over (2\pi)^3}\left(\sqrt{{\bf k}^2+m_B^2}-\sqrt{{\bf k}^2
+m_F^2}\right) \, \simeq \, N\frac{m_B^2\Lambda^2}{8\pi^2}.
\ee
As we will see later, due to this quantum correction to the cosmological
constant,  some
soft parameters are changed by an amount of
order $N {\alpha_{\rm GUT}}/{2\pi}$ times the tree level values.
Then a simple but important point is that although
$\alpha_{\rm GUT}/2\pi$ is small, $N \alpha_{\rm GUT}/{2\pi}$
can be significantly larger in realistic models.
In this paper, we wish to discuss  this point
together with its implications
in the context of a simple supergravity model.

To proceed, let us consider a simple supergravity model with the
following K$\ddot {\rm a}$hler potential and
superpotential\cite{nilles}
\begin{eqnarray}
K&=&h_{\alpha}h^*_{\alpha}+\phi_i\phi^*_i
+(\xi H_1H_2+{\rm h.c.})\ ,
\nonumber \\
W&=&W_h(h_{\alpha})+\frac{1}{6}
{\lambda}_{ijk}\phi_i\phi_j\phi_k,
\end{eqnarray}
where $h_{\alpha}$ denote hidden sector
fields triggering SUSY breaking, and
$\phi_i$ are generic observable matter fields
including quarks, leptons, and the two Higgs doublets $H_1$ and $H_2$.
The corresponding supergravity lagrangian can be expanded in
terms of $\phi_i$
as
\be
{\cal L}={\cal L}_h-
(m^2_{3/2}+\kappa_0^2 V_h)\phi_i\phi_i^*
 -[\xi (\kappa_0^2 V_h+2 m^2_{3/2})
H_1H_2+
{\rm h.c.}]+...,
\ee
where ${\cal L}_h\equiv {\cal L}(\phi_i=0)$,
and the ellipsis stands for other observable-field-dependent terms.
Here the local composite operators
$m_{3/2}^2$ (the gravitino mass operator)  and $V_h$ (the hidden sector
scalar potential operator) are given by
\begin{eqnarray}
&&m_{3/2}^2=\kappa_0^4 |W_h|^2\exp (\kappa_0^2 h_{\alpha} h^*_{\alpha}),
\nonumber \\
&&V_h= (|D_{\alpha} W_h|^2-3\kappa_0^2 |W_h|^2)\exp(\kappa_0^2 h_{\alpha}
h^*_{\alpha}),
\end{eqnarray}
where
$D_{\alpha}W_h=(\partial_{h_{\alpha}}+\kappa_0^2 h^*_{\alpha})W_h$.

For a given supergravity lagrangian ${\cal L}$, one may integrate
out the hidden fields to
obtain the effective lagrangian ${\cal L}_{\phi}$
of the observable fields $\phi_i$:
\be
\exp (i \int d^4x \, {\cal L}_{\phi}) = \int [{\cal D} h] \exp (i
\int d^4x \sqrt{g}\, {\cal L}).
\ee
Here $[{\cal D} h]$ denotes the integration of the
hidden fields including the gravity
multiplet  $(g_{\mu\nu}, \psi_{\mu})$.
If the vacuum values of $W_h$ and $D_{\alpha}W_h$
are nonzero and thus SUSY is broken, ${\cal L}_{\phi}$
would contain the soft SUSY breaking terms:
\be
{\cal L}_{\phi} \, \ni \, -\ m_0^2\phi_i\phi^*_i
-(\frac{1}{6}A_{ijk}\phi_i\phi_j\phi_k+BH_1H_2+
{m}_a\lambda^a\lambda^a+\rm h.c.),
\ee
where $\lambda^a$ are the gauginos.

Since none of the $\phi_i$-modes are
integrated out, the effective lagrangian ${\cal L}_{\phi}$ is essentially
renormalized at the cutoff
scale $\Lambda$,  viz  all the operators and the parameters
in ${\cal L}_{\phi}$ are renormalized at $\Lambda$
in the sense of Wilson.
Among the coefficients of soft terms,
$m_0^2$ and $B$  depend on the expectation value
of the hidden sector scalar potential, $\langle V_h\rangle$,
while the others are independent of $\langle V_h\rangle$.
 From  Eqs. (3) and (5), one easily finds
\be
m_0^2(\Lambda)=\langle m^2_{3/2}+\kappa_0^2 V_h\rangle,
\quad
B(\Lambda)=\langle \xi(\kappa_0^2 V_h+2m_{3/2}^2)\rangle,
\ee
where the bracket means the average over the hidden
fields, e.g.
\be
\langle V_h\rangle =\int [{\cal D} h] V_h(h_{\alpha}) \exp (iS_h)/
\int [{\cal D} h] \exp (i S_h),
\ee
where $S_h=\int d^4x \sqrt{g} \, {\cal L}_h$.

Obviously
both $\langle m^2_{3/2}\rangle$ and
$\langle V_h\rangle$
are determined entirely by the hidden sector parameters.
To be  phenomenologically acceptable,
those hidden sector parameters must be adjusted to the values
leading to the vanishing cosmological constant.
In most of the previous studies of soft terms, motivated by
the vanishing cosmological constant,
$\langle V_h\rangle$ has been taken to
be zero or  $\kappa_0^2\langle V_h\rangle \ll m_{3/2}^2$.

Does the vanishing cosmological constant
imply $\kappa_0^2\langle V_h\rangle \ll m_{3/2}^2$ ?
The cosmological constant
$V_{\rm eff}$ can be obtained by
integrating out all fields in the theory$\cite{weinberg}$:
\be
\exp (i \int d^4x \, V_{\rm eff})=
\int [{\cal D}\phi {\cal D} h] \exp (i \int d^4x \sqrt{g}
\, {\cal L}),
\ee
where $[{\cal D}\phi]$ stands for the integration
of all observable gauge and matter multiplets.
Clearly {\it at tree approximation} where
the path integral
representations  of
$\langle V_h\rangle$
and $V_{\rm eff}$ (see Eqs. (8) and (9))
are saturated by the field configurations satisfying the classical
equations of motion,
we have
$(\langle V_h\rangle)_{\rm tree}=(V_{\rm eff})_{\rm tree}$, and thus
the hidden sector parameter values
yielding a vanishing tree level cosmological constant
automatically gives $(\langle V_h\rangle)_{\rm tree}=0$.
However, as can be noticed by their definitions,
$V_{\rm eff}$ and
$\langle V_h\rangle$ can receive completely different quantum
corrections.

Even at one loop order, quantum fluctuations
of both the observable fields  and the hidden fields
give a quadratically divergent correction
to the cosmological constant$\cite{zumino}$:
\be
\delta V_{\rm eff}\equiv
V_{\rm eff}-(V_{\rm eff})_{\rm tree} \, \simeq \,
\frac{1}{8\pi^2}{\rm Str} \,  ({\cal M}^2) \, \Lambda^2,
\ee
where ${\cal M}^2$ denotes the mass matrix of the supergravity
action.
For our simple model, we have
\be
{\rm Str} \, ({\cal M}^2) \,\simeq\, (Nm_0^2-\tilde{N}\tilde{m}^2)
+... \equiv \,
N_{\rm eff}m_0^2(\Lambda),
\ee
where $N$ is the number of observable chiral multiplets,
$\tilde{N}$ is the number of gauginos which
are assumed to have a common mass $\tilde{m}$, and
the ellipsis denotes the contribution
>from the hidden sector.
Here we have neglected the masses of matter fermions and gauge
bosons.
For the hidden sector contribution,
the gravity multiplet
gives a negative contribution  ($=-4m_{3/2}^2$) while
the hidden chiral multiplets give a positive contribution
proportional to the number of multiplets.

Unlike the cosmological constant,
$\langle V_h\rangle$ receives a correction only from the hidden
field fluctuations.
For $\kappa_0\Lambda=g_{\rm GUT}$,
if (i) interactions among hidden sector fields are weak enough,
and (ii) the number of hidden multiplets
which contribute to SUSY breaking by having nonvanishing
auxiliary components
is of order unity,
we have
\be
\delta\langle V_h\rangle =O(\frac{\alpha_{\rm GUT}}
{\pi}|F_{\alpha}|^2)=O(\frac{\alpha_{\rm GUT}}{\pi}\kappa_0^{-2}
m_{3/2}^2),
\ee
where $\delta\langle V_h\rangle = \langle V_h\rangle -(\langle
V_h\rangle )_{\rm tree}$, and the $F$-term of $h_\alpha$ is
given by
$F_{\alpha}=D_{\alpha}W_h \exp (\kappa_0^2h_{\beta}h^*_{\beta}/2)=
O(\kappa_0^{-2}m_{3/2}^2)$.
In fact, the conditions (i) and
(ii) are met by many simple hidden sector models.
One might think that these conditions
are not fulfilled by the popular gaugino condensation
model$\cite{gcondensation}$ for SUSY breaking in string theory.
However, in the gaugino condensation model,
strongly interacting gauge nonsinglet hidden multiplets can be
integrated out without breaking SUSY.
Then the effects of integrating out gauge nonsinglet
fields  can be summarized by
an effective superpotential of
a relatively small number of weakly interacting
gauge singlet multiplets, e.g. the dilaton and moduli
multiplets, for which the conditions (i) and (ii) are fulfilled.

Using Eqs. (7), (10), and (12) together with $(V_{\rm eff})_{\rm tree}=
(\langle V_h\rangle)_{\rm tree}$, we easily find that
the phenomelogical requirement of $V_{\rm eff}=0$ leads to
\be
\kappa_0^2\langle V_h\rangle\simeq
-\epsilon m_0^2(\Lambda)\simeq -\frac{\epsilon}{1+\epsilon}
\langle m_{3/2}^2\rangle,
\ee
where
\be
\epsilon=\frac{N_{\rm eff}}{8\pi^2}(\kappa_0\Lambda)^2.
\ee
In many cases, the gaugino mass contribution to $N_{\rm eff}$
is expected to be significantly smaller than the
chiral matter contribution.
It is then quite conceivable
that
$N_{\rm eff}$ is {\it positive} and of $O(8\pi^2)$,
implying that
$\epsilon$ can be essentially of order unity for
 $\kappa_0\Lambda=g_{\rm GUT}$.
Note that $N_{\rm eff}$ receives a contribution
>from all chiral multiplets with masses far below
$M_{Pl}$, particularly from those in the minimal supersymmetric standard
model containing 49 chiral multiplets.

So far, we have considered only
the quadratically divergent one loop contribution
to the cosmological constant, which is
of order $\kappa_0^{-2} m_{3/2}^2$ due to a
large value of $N_{\rm eff}$ compensating over
the loop
suppression factor $1/8\pi^2$.
Higher loops also give quadratically
divergent contributions to the cosmological constant.
However, contrary to the one loop effect
higher loop effects do {\it not} contain any
additional large factor
which may compensate the additional loop suppression
factor $1/8\pi^2$.
For example, two loop diagrams involving gauge interactions
give a contribution smaller than the one loop
effect by the small factor $\alpha_{\rm GUT}/2\pi$.
For two loop diagrams
involving trilinear and/or Yukawa interactions also, it
is suppressed by (coupling constant)$^2$.  For the trilinear
couplings, the two loop diagram is shown in Fig. 1 and its
contribution is roughly
$\frac{1}{(4\pi^2)^2}\sum_{ijk}A_{ijk}A^*_{ijk}
\Lambda^2$.  For Yukawa interactions, the contribution
is roughly
$\frac{1}{(4\pi^2)^2}\sum_{ijk}\lambda_{ijk}\lambda^*_{ijk}
m_{3/2}^2\Lambda^2$.
Usually $A_{ijk}$ is of order $m_{3/2}\lambda_{ijk}$,
and then clearly these two loop effects are negligible
compared to the one loop effect of Eq. (10).
Then our result of Eq. (13) which was derived within
the one-loop approximation
would remain valid even after taking into account
higher loop effects.
One may also include the effects associated with the electroweak
symmetry breaking and the nonperturbative QCD effects.
Since  $\Lambda^4_{\rm QCD}\ll m_W^4\ll
\kappa_0^{-2} m_{3/2}^2$, again
this point does not affect at all  our result of Eq. (13).

In the above discussion, we have pointed out that, if one implements
a fine tuning
of the hidden sector parameters
to have the vanishing {\it tree level}
cosmological constant, the corresponding hidden sector
scalar potential $V_h$ has an expectation value
$|\langle V_h\rangle|\ll \kappa_0^{-2} m_{3/2}^2$, which is the relation
frequently used in the previous analyses of soft parameters.
However, if the hidden sector
parameters are adjusted to yield the vanishing
{\it renormalized} cosmological constant, $\langle V_h\rangle$
is likely to have
a {\it negative} value of $O(\kappa_0^{-2}m_{3/2}^2)$.
Note that this is not due to a quantum correction
of $\langle V_h\rangle$ in the path integral
evaluation,
which is of $O((\alpha_{\rm GUT}/\pi)\kappa_0^{-2}m^2_{3/2})$
if the two plausible conditions specified above Eq. (12) are met,
but due to a change in the fine tuning of the hidden sector
parameters  required for the vanishing renormalized cosmological
constant.

What would be the implications of our observation
that $\kappa_0^2\langle V_h\rangle=-O(m^2_{3/2})$?
First of all, it implies that
the $\langle V_h\rangle$-dependent parts
of soft parameters can be sizable, and thus must
be included when one computes soft parameters for a given
supergravity model, as was done recently
by Brignole, Ibanez and Munoz$\cite{ibanez}$.
In some cases, using the input $\langle V_h\rangle=0$,
one obtains a very simple pattern of soft parameters
renormalized at $\Lambda$.
For instance,
in the dilaton-dominated SUSY breaking scenario in string theory,
if the $\mu$-term is induced entirely by the K\"{a}hler potential
term $\xi H_1H_2$\cite{masiero},
one finds the relation
$m_0:B/\mu:A_{ijk}/\lambda_{ijk}=1:2:-\sqrt{3}$$\cite{ibanez,louis}$.
We have already noted that $m_0^2(\Lambda)$ and $B(\Lambda)$
associated with $\xi H_1H_2$ contain
$\langle V_h\rangle$-dependent pieces, while the other
soft parameters are independent of $\langle V_h\rangle$.
Clearly then, for $\kappa_0^2\langle V_h\rangle=-O(m_{3/2}^2)$,
the above relation must be modified
and only the property of soft parameters which is independent
of the input $\langle V_h\rangle=0$ must be seriously
taken into account$\cite{ibanez}$.

There is another implication of our observation.
It has been pointed out that
in multi-gaugino condensation models$\cite{racetrack}$
for SUSY breaking in string theory,
the hidden sector potential, i.e. the dilaton potential,
appears to have a negative value of $O(\kappa_0^{-2}m_{3/2}^2)$
for a reasonable
range of the hidden sector parameters.
This negative dilaton potential has been often considered
as an undesirable feature of the model.
However, our discussion above indicates that
a negative value of the hidden sector potential
is not a problem, but rather a desirable feature,
for the fully renormalized cosmological constant to vanish.

A few comments should be made.
To obtain the soft parameters renormalized at
$m_W$ from those at $\Lambda$,
one must integrate out the $\phi_i$-modes with frequencies
greater than $m_W$.
This would lead to a subsequent renormalization
of soft parameters.
Such  subsequent remormalizations,
e.g. $m^2_0(\Lambda)/m^2_0(m_W)$,
are determined mainly
by the observable sector couplings in ${\cal L}_{\phi}$,
e.g. the gauge and Yukawa couplings,
and thus are {\it independent} of
the one-loop renormalization of the cosmological constant
which affects only the hidden sector parameters.
Note that both $(V_{\rm eff})_{\rm tree}$ and
the one-loop correction $\delta V_{\rm eff}$ of Eq. (10)
depend only on the hidden sector parameters.

Since the quantum corrections to the cosmological constant of order
$\kappa_0^{-2}m_{3/2}^2\gg m_W^4$ are added to the tree level value,
one might wonder whether
the vacuum structure of the observable sector fields $\phi_i$
is changed.
However, due to SUSY,
the effective potential of $\phi_i$ does not contain
any field-dependent quadratic
divergence. As a result,
the corresponding vacuum structure
is {\it not} touched at all by the quadratically divergent correction to
the cosmological constant.
Also our results of Eqs. (13)
explicitly show that
$m_0^2(\Lambda)$ is still {\it positive}
even after including the contribution
>from a {\it negative} $\langle V_h\rangle$.

In conclusion, we have pointed out in this paper that
the quantum correction to the cosmological
constant in supergravity models is likely to be {\it positive}
and  can be of $O(\kappa_0^{-2}m_{3/2}^2)$, while the
quantum correction to $\langle V_h\rangle$ is of
$O(\pi^{-1}\alpha_{\rm GUT}\kappa_0^{-2}m^2_{3/2})$.
This is for a reasonable  choice of the cutoff scale,
$\kappa_0\Lambda=g_{\rm GUT}$, and mainly due to
a large number of  chiral multiplets  conpensating over
the loop suppression factor.
Although they are equal
at tree level, the renormalized
cosmological constant and the expectation value of
the hidden sector scalar potential  $V_h$ can differ from each other
by an amount of $O(\kappa_0^{-2}m_{3/2}^2)$.
Then, the condition of the vanishing cosmological constant
leads to a {\it negative} $\kappa_0^2\langle V_h\rangle$ of
$O(m_{3/2}^2)$. This means
the $\langle V_h\rangle$-dependent parts of the soft parameters
cannot be ignored, and thus the conventional studies
of supergravity phenomenology
based on the input $\kappa_0^{2}\langle
V_h\rangle \ll m_{3/2}^2$
should be
accordingly modified.
This also implies that it might
not be a problem, but  rather a desirable feature
(for the fully renormalized cosmological constant to vanish)
that the hidden sector scalar potential has a negative
minimum whose magnitude is of
$O(\kappa_0^{-2}m_{3/2}^2)$.

\acknowledgments
This work is supported in part by Korea Science and Engineering
Foundation through Center for Theoretical Physics at
Seoul National University (KC, JEK),  KOSEF--DFG Collaboration
Program (JEK, HPN), the Ministry of Education of The Republic of
Korea (KC, JEK), Deutsche Forschungsgemeinschaft (HPN) and EC
grants SC1-CT91-0729 and SC1-CT92-0789 (HPN).

\begin{figure}
\caption{A two loop contribution to the vacuum energy from the $A$
terms.}
\end{figure}

\end{document}